University of Warsaw
Faculty of Economic Sciences

Grzegorz Krochmal

# Sentiment of tweets and socio-economic characteristics as the determinants of voting behavior at the regional level. Case study of 2019 Polish parliamentary election.

Master degree thesis
Field of the study: Data Science and Business Analytics

Warsaw, July 2020

# Summary

This work is dedicated to finding the determinants of voting behavior in Poland at the poviat level. 2019 parliamentary election has been analyzed and an attempt to explain vote share for the winning party – Law and Justice has been made. Sentiment analysis of tweets in Polish (original) and English (machine-translations), collected in the period around the election, has been applied. Amid multiple machine learning approaches tested, the best classification accuracy has been achieved by Huggingface BERT on machine-translated tweets. OLS regression, with sentiment of tweets and selected socio-economic features as independent variables, has been utilized to explain Law and Justice vote share in poviats. Sentiment of tweets has been found to be a significant predictor, as stipulated by the literature of the field.

# Key words

sentiment analysis, Twitter, tweets, machine learning, neural networks, voting behavior, elections

# Field of the thesis (codes according to the Erasmus program)

Economics (14300)

**TABLE OF CONTENTS**

# INTRODUCTION

In the modern world, especially in well developed countries internet access is perceived as the basic human right. People can use it whenever and wherever they want to. Statistics prove that they do it willingly, with 1.7 megabytes of data generated per second by every person on earth (DOMO, 2018). Not all of people have equal access to the internet. In 2017 the number of individuals able to connect to the global network was up to 3.8 billion (DOMO, 2018). It can be then estimated that every internet user generates about 300 megabytes of data per day. This is not just an empty number. The data we generate says a lot about the way we live and who we are. According to the studies conducted by Kosinski, Stilwell and Graepel (2013), based on likes people leave on Facebook, many private information might be guessed by an algorithm with high accuracy. The detailed performance of the model derived by the scientists can be seen in Figure 1.

*Figure 1. Classification accuracy of Kosinski et. al (2013) algorithm based on Facebook likes*

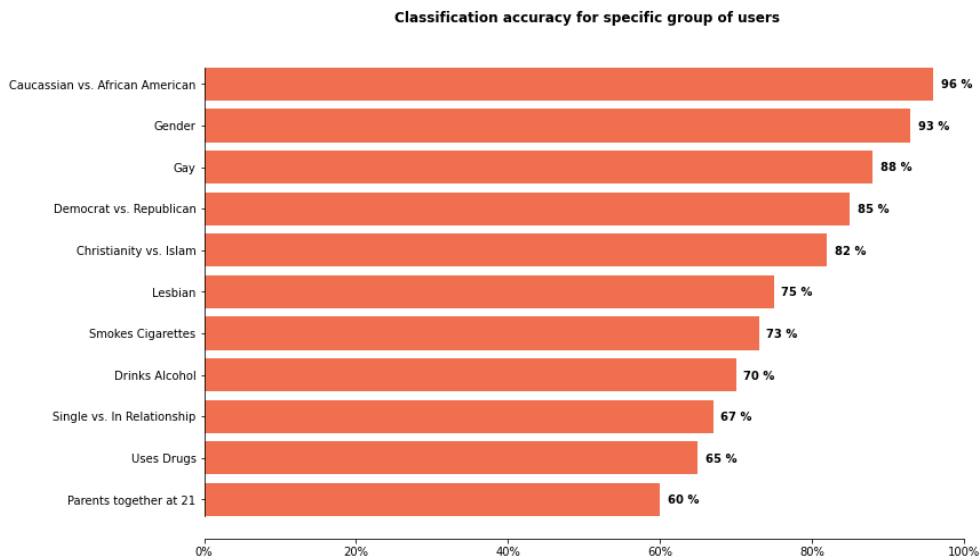

What is worth mentioning, with respect to the presented results, is that the study utilized only the information people share voluntary. This means they were more or less aware that their behavior is not anonymous. However, this consciousness of users might not be the case for other online activities i.e. online shopping, maps, google search and many others, which are constantly recorded and can be used by vast number of institutions to get deepened knowledge of the individual.

The issue when people find out that the data they generate have many, sometimes shady, applications got a serious attention after 2016 United States presidential election and Brexit. The media reported that the results of both of the campaigns were influenced by Cambridge Analytica. The company exploited information people leave on social medias to profile their political preferences. The campaigners who hired Cambridge Analytica were able to deliver more personalized content to the voters (Rosenberg et al., 2018). However, the history of taking advantage of the internet and social medias to influence elections is not limited to Trump's successful presidential campaign or Brexit. Guriev, Melnikov and Zhuravskaya (2019) analyzed data about 3G mobile network coverage in Europe and its link to voting behavior. The authors reported significant rise in vote share of populist parties[1] in places with better accessibility to mobile internet networks. In other study, led by Campante, Durante and Sobbrio (2013), the researchers showed that the evolution of social medias opened gates for citizens to get more easily involved in protests and antiestablishment movements. This, in the effect, brought growth of popularity of political parties supporting such initiatives. The authors, to strengthen the statement, gave an example of Five Star Movement. The party started as a marginal political fraction, led by a blogger and comedian, to become the important member of Italian political scene a few years later. They won the election in 2018 and formed the government in coalition with Lega Nord. The rise of the party's popularity was the most noticeable in places with better access to the broadband internet.

The main aim of this work is to combine the knowledge which can be extracted from social medias with socio-economic data to explain the poviat[2]-level results of Polish parliamentary election, which took place on the 13th of October 2019. Based on the relevant literature on the subject and the issues identified, two main hypotheses are proposed.

*Hypothesis 1*:

*Results of the elections at the poviat level can be explained with socio-economic variables, as suggested by Fielding (1998).*

---

[1] But only these in opposition.
[2] "Powiat (or poviat) is the second-level unit of local government and administration in Poland" https://en.wikipedia.org/wiki/Powiat

*Hypothesis 2:*

*Sentiment of tweets, aggregated at the poviat level, is a statistically significant predictor of election outcomes.*

To make their verification possible, two separate research problems need to be addressed in this study.

First, it is essential to obtain valuable information from social medias, transform them and afterwards model along with socio-economic variables. Social media platform chosen for this study is Twitter. To encode, for further modeling, the collected data, sentiment analysis on tweets is introduced. This is the method which allows to extract an emotional state of specific text.[3] Sentiment analysis is present in many fields varying from business applications to social studies. The tweets collected for the purpose of this work are in Polish. This might be considered as a complication, because the vast majority of tools available in the field have been created for English language. To make it possible to label the tweets with sentiment scores, differentiated machine learning approaches are tested. Trying out distinct strategies can provide not only the best possible classification accuracy but also increase the understanding of pros and cons of each of the methods.

Second, when the best performing sentiment model is obtained, collected tweets are labeled with the sentiment scores. The data have been gathered with location details. This allows to obtain poviat-level sentiment results. In this study, they are perceived as a regional measure of social happiness.

Regional sentiment is combined with socio-economic data and modelled together to verify research hypotheses. The task is completed with OLS regression. This algorithm has been selected, due to its interpretability, which makes the hypotheses verification directly addressable.

The work is divided into four sections. Section 1 contains the literature review. At the end of the section, literature-derived hypotheses are proposed. In Section 2 the whole process of sentiment analysis conducted on tweets is described. Section 3 includes OLS regression with poviat-level sentiment scores and socio-economic variables as descriptive features. In Section 4 the final results and discussion can be found.

---

[3] In most applications negative/neutral/positive classification is availed

# SECTION I

## Literature overview

The internet has made unprecedented changes in the way people communicate. With the rise of social media platforms it has never been easier to connect with others. Nowadays, everyone with access to the internet can share whatever with whoever they want to. The most popular sites experience enormous traffic with hundreds of millions of active users everyday. The popularity of some of the biggest platforms is visualized on Figure 2.

*Figure 2. Monthly active users of most popular social media platforms*

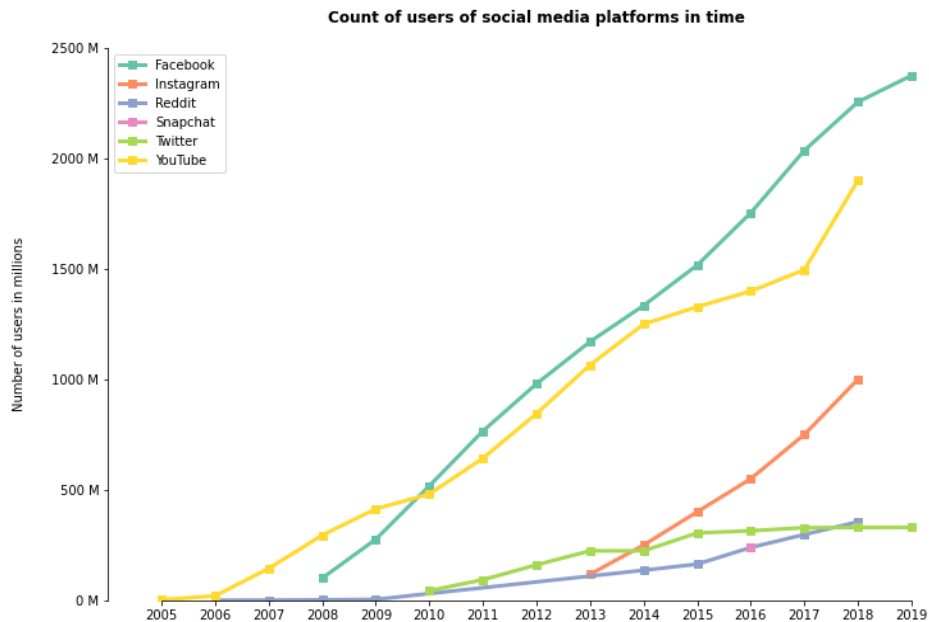

*Source:* Own elaboration on the basis of data obtained from https://ourworldindata.org/rise-of-social-media (Ortiz-Ospina, 2019)

Such a phenomenon, where people voluntary share information about themselves and interact with each other on a huge scale, made the social media platforms a vast and unique data source for different social researches. This work focuses on Twitter. Therefore, the literature overview of potential applications of social medias in understanding political outcomes is limited to this platform.

Twitter is a micro-blogging site, where the basis of communication between users are short texts with the limit of 280 characters. There are no restrictions on topics which can be covered by bloggers. Tweets can be analyzed in order to get a wide and easily accessible understanding of differentiated social issues. The relation between tweets and politics is a broadly studied topic, especially since Barack Obama's presidential victory in 2008. It was the first campaign, during which a politician used social medias on a large scale to communicate with voters. It is enough to say that when Obama managed to get 5 million online supporters, his opponent – McCain had a little bit more than 1 million of fans. On Twitter, the dominance of Obama was even clearer as he collected 115k followers, with McCain having 23 times less (Chang & Aaker, 2010). In less than 2 years Obama underwent a change from a not really known senator from Iowa to the first African American president. That campaign was the clear signal that the significance of social media in case of politics cannot be underestimated. Experts started to perceive this source as the tool both to influence elections and to understand their results.

In 2010 the group of German researchers analyzed tweets published in a few weeks which preceded 2009 parliamentary election in Germany. One of the main conclusions presented in the article was that "the mere number of tweets mentioning a political party can be considered a plausible reflection of the vote share and its predictive power even comes close to the traditional election polls" (Tumasjan et al., 2010). In other words, it was stated that it is enough to asses the share of tweets mentioning particular party in a group of all political tweets to predict results close to the real-life vote shares. Such a discovery attracted the attention of many researchers. Similar conclusions were obtained by Sang & Bos (2012). The authors analyzed tweets coming from a two-week period before Dutch senate elections. They used the method suggested by Tumasjan et al., (2010) and compared the results obtained with polls[4] to conclude that the predictions seem to be accurate. However, the authors were not as confident as Tumasjan et al., highlighting that they failed to measure statistical significance of the results. Additionally, Sang & Bos pointed out the issue with the demographic structure of Twitter users, reporting that it can be skewed into a direction of younger people, even these who cannot vote yet. At the same time, the role of older voters, who, in general, are more engaged in voting process, is probably marginalized. Tumasjan et al., study was replicated also in the case of 2011 Singapore general election (Skoric et al., 2012). Singapore's political system differs from fully democratic German or Dutch structures. The authors stressed that when "old type" media, like press or tv, are dominated by the ruling party – People's Action Party, social medias are the place where oppositional movements focus their actions. That was raised as a possible bias source. Nevertheless, the estimations occurred to be consistent with these obtained by Tumasjan et al. However, the authors reported that the mean absolute error of vote share predictions was significantly higher – 5.23 % in comparison to 1.65 % obtained by German researchers. Again, as Sang & Bos argued, Skoric et al., pointed out that, even if Tumasjan et al., approach provides somehow correct results, it does not incorporate many important features of Twitter, such as characteristics of users. Furthermore, the authors suggested that the approach might be improved by incorporation of sentiment analysis. Eventually, it is rather obvious that a political party can be mentioned with a negative attitude and an author of a post is not going to vote for the party it writes about. That is not captured by simple Tumasjan et al., methodology.

---

[4] As opposed to Tumasjan et al., who chose real-life results as a benchmark.

Dutch and Singapore-based studies, which utilized Tumasjan et al., methodology, raised a question whether Twitter's population is representative enough to consider it as a basis for nationwide studies in the field of political choices. That issue was addressed by Gayo-Avello et al., (2011). In 2011 the researchers tested Tumasjan et al., method on tweets regarding 2010 United States Senate special election in Massachusetts and United States Congressional election. They confronted outcomes of six different races – two political opponents fighting for one position – with the predicted results based on share of tweets regarding each candidate. The predictions obtained were accurate only for three of the cases. Having in mind that only six races were analyzed, the accuracy of classification would be the same for random classifier (Gayo-Avello et al., 2011). Gayo-Avello conducted also another study in the attempt to explain results of 2008 United States presidential election. This time the author tested a few approaches – already described Tumasjan et al., methodology and two others, based on lexicon type sentiment analysis. For the latter two he checked the sentiment of tweets, which contained the name of the candidates. The author assumed that people would vote for the candidate with the most positive sentiment score of all tweets in which specific politician was mentioned. Neither of the methods tested provided predictions which would be able to outperform random classifier. Gayo-Avello highlighted the issue with population representability for Twitter data. He argued that it is highly likely that people who share their political views on Twitter are the specific subgroup and cannot be treated as the global population (Gayo-Avello, 2011).

Tumasjan et al., method applied to different political data turned out to be, in some way, accurate and able to capture general tendencies. However, due to representativity issues and instability, it is not sufficient to replace standard polls. To tackle mentioned limitations, researchers proposed other strategies. DiGrazia et al., (2013) combined Tumasjan et al., approach with district level socio-economic features to analyze the 2010 and 2012 United States electoral elections. The authors used OLS regression to measure influence of selected variables on real-life vote share for Republicans and Democrats. DiGrazia et al., stated that share of tweets, mentioning particular party, occurred to be statistically significant predictor. This observation coincides with Tumasjan et al. Socio-economic variables did not have significant impact on the outcomes of elections.

Ceron et al., (2014) made an attempt to predict political vote shares with sentiment analysis. The researchers built their analysis upon HK method (Hopkins & King, 2010) which allows to asses the percentage sentiment structure of some set of documents– to what extent they are positive/negative. The authors utilized tweets to predict outcomes of 2012 French legislative election. They reported mean average error of their predictions to be equal to 2.38% which, according to researchers: "(…) is not far from the MAE values displayed by the surveys held in the last week before the elections. On average, survey polls registered MAE equal to 1.23% (…)". The comparison of predictions obtained by authors with real-life results can be seen in Figure 3.

*Figure 3. Predicted and actual vote shares related to the first round of the 2012 French elections.*

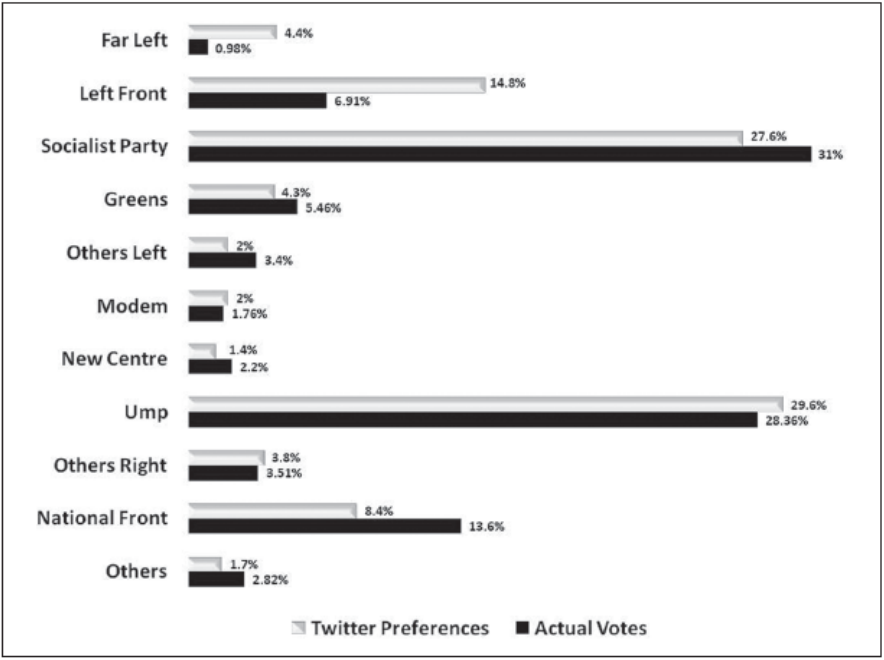

*Source:* Ceron et al. (2014)

The results presented in Figure 3 seem to be quite accurate. However, there are some issues with this paper. The authors did not precisely specify how they had calculated presented scores.

The HK method is a 7 level classification problem: "extremely negative (−2), negative (−1), neutral (0), positive (1), extremely positive (2), no opinion (NA), and not a blog (NB)[5]" (Hopkins & King, 2010). It can be deducted, based on another example from the Ceron's paper, that the score is obtained as a positive to negative ratio of tweets, but this is not sufficient knowledge, taking into account that classes returned by HK method are not limited to a binary – – negative/positive split. Another issue is that Ceron et al., were surprisingly picky in a description of possible sources of badly predicted cases. While they explained that National Front misclassification probably came from online underrepresentation of people sympathizing with the party, they did not said a word about Left Front, for which the error was even higher. Additionally, the authors did not deliver any information about the key according to which tweets were selected and how they were preprocessed. Ceron et al., paper is interesting and to some level proves that sentiment analysis might be an useful tool for the prediction of political outcomes. At the same time, there is some haziness in the case of the methodology. This issue makes the work not really replicable and also not fully reliable.

The attempt to use sentiment analysis in explaining the results of elections was also made in the case of 2013 Bulgarian parliamentary election (Smailović et al., 2015). The authors chose a machine learning approach to classify political tweets which were posted before the voting. To have a training dataset, they collected a larger set of general tweets and engaged a group of qualified people to annotate texts as negative, neutral or positive. Twenty-nine thousand tweets were manually labeled. Researchers trained SVM model on the annotated data and classified ten thousand political tweets to get their sentiment. When general tweets – manually annotated, were balanced in case of percentage count – with 42 % neutral, 33 % positive and 25% negative, the political tweets, classified by the model derived by researchers, consisted mostly of negative cases. The authors reported MAE to be equal to 2.09 %. For the comparison they also tested Tumasjan et al., (2010) vote share approach, which achieved 1.88 % MAE. The sentiment-based predictions of election scores in the case of mean average error look promising. However, when studied deeper, they are counterintuitive. According to the obtained scores, the party with the higher share of negative tweets is more likely to win the election. It is not the real-life tendency, but just the effect of imbalance of predicted sentiment labels, with negative tweets outnumbering positive.

A few applications of Twitter data to the modelling of political outcomes have been discussed in previous paragraphs. However, already presented examples do not fully relate to the goal of this work. Sentiment analysis in this study is conducted on the regional level in order to obtain the measure of happiness for particular poviats in Poland. The measure is then combined with the socio-economic data in the attempt to explain the results of the elections at the regional level. There is then one last question which has to be addressed. What are the socio-economic determinants of political choices? A wide study of this issue was conducted by Fielding (1998). The author analyzed 1992 general election in Scotland. He suggested the following determinants of voter's choice at the constituency level:

- The unemployment rate
- Average current disposable household income
- Average household wealth
- The proportion of the population holding a higher educational qualification
- The proportion of the population employed in occupational classes 1 and 2[6]
- The proportion of the working population employed in agriculture.

---

[5] Hopkins & King work was focused on analysis of different blogs – this is not the case of Ceron et al. study conducted on Twitter.

Currently known achievements in employing Twitter as the descriptor of election outcomes have been discussed. Presented literature, however wide and valuable, does not close the subject. No studies were found, where information about the sentiment from tweets was combined with socio-economic variables to understand voting behavior at the regional level. Lack of such materials remains true also for the attempt to employ Twitter to explain political choices in Poland. Therefore, this thesis might bring an additional value to the topic. To relate to literature and properly place the research problem, the following two hypotheses are proposed:

*Hypothesis 1:*

*Results of the elections at the poviat level can be explained with socio-economic variables, as suggested by Fielding (1998).*

*Hypothesis 2:*

*Sentiment of tweets, aggregated at the poviat level, is a statistically significant predictor of election outcomes.*

---

[6] „This is a measure of the size of the middle class population as a fraction of the total. This variable reflects the impact of social class on political attitudes, controlling for income and wealth. " (Fielding, 1998)

The hypothesis are verified with machine learning and econometric methods introduced in Section 2 and used in Section 3.

# SECTION II

## Sentiment analysis of tweets

On the day this paragraph is written, 1$^{st}$ April 2020, the whole world is paralyzed by the spread of a potentially deadly disease. The COVID-19 pandemic has completely changed the reality people in developed countries were used to. Day after day, new restrictions, introduced by governments, took away many of the privileges which, up to these days, had seemed to be obvious. In Poland, restaurants, cinemas, schools have all been closed for more than 2 weeks now. Everyone, if it is possible, works at home. Three months ago, when first news about an unknown virus killing citizens of Wuhan appeared in media, not many people in Western countries considered this as a real threat. The governments were not introducing any kinds of special prevention measures, living a dream that China is able to contain the pathogen in its borders. Now it is obvious that this was just a wishful thinking. The neglection is even more surprising, knowing that it is not probably the case the Westerners found out about the threat too late. Actually, Canadian algorithm – BlueDot was alarming about Wuhan's disease before it was reported to the wider audience by China. It had sent first alert nine days before WHO released its statement (Stieg, 2020). This striking example of the algorithm being more effective in the response to the real-life problem than the reputable institution has a connection with the idea of this work.

BlueDot algorithm bases on big data gathered all around the web with high frequency. It combines natural language processing and machine learning tools to collect valuable information. The idea of social media mining and sentiment analysis, which is an important part of this study, is then similar to the way BlueDot works. Sentiment analysis enables automatic detection of affective states in texts. It is easy to imagine, especially in the age of social medias, where there is easily accessible data almost everywhere, that the method has almost unlimited variety of possible applications, e.g.:

- Hate speech detection (MacAvaney et al., 2019; Petrocchi & Tesconi, 2017)
- Marketing research (Rambocas & Gama, 2013)
- National security (Earhart, n.d.).

Obviously, there are much more study fields where sentiment analysis can be applied, among which is the subject of this work – politics and regional differentiation.

There are several issues which are addressed in this section. First, an overview of data – collected tweets. Next, the general strategy to the modelling is outlined. Text preprocessing steps and the machine translation approach is described. Thirdly, there is a part devoted to the sentiment analysis modelling with all the methods discussed and their performance presented. At the end, the sentiment scores are predicted for the collected tweets and the regional differentiation in happiness is analyzed.

## 2.1. Data overview

The parliamentary elections in Poland took place on 13$^{th}$ October 2019. The tweets for this study were collected from 14$^{th}$ September – 14$^{th}$ November with the Twitter API[7] and Python wrapper for it – Tweepy[8]. The data therefore covers the period one month before and after the election. The analysis in this study is conducted at the poviat level. There was then a need to leave only the tweets with a specified user's location. Other languages than Polish were not considered.

At the initial stage there were 236,374 tweets in the database. The first issue, which had to be addressed was that Twitter API as a location returns only the name of a city. The location-level which is used in this study is a poviat. Adjustment from cities to poviats had to be done. This process was sequential and is precisely described in Annex 1. At the end of the process, the tweets contained the full location data: voivodeship, poviat, commune and geocoordinates. Afterwards, text preprocessing operations were introduced. This step is deeply described in Section 2.2.2. The final, fully preprocessed dataset, consisted of 122,719 tweets.

Before moving to further subsections, it is essential to discuss the limitations of the data. There are two potential problems which can be easily spotted:

- Time frames – the data was collected before and after the event which was the main concern area of the study – 2019 Polish parliamentary election.
- Population representativeness.

First, there might arise the question whether the joint analysis of time-series data, where the event which is analyzed happens in the central period, is a correct approach. If the tweets and their sentiment, which is studied in the Section 2.3, are the reflection of regional happiness, what if that happiness was influenced by the results of the election. In other words what if the event, which is the subject of study, influenced the data, which is the basis for the explanation

---

[7] https://developer.twitter.com/en/docs/tweets/search/overview
[8] http://docs.tweepy.org/en/v3.5.0/

task. To weigh whether this might be a threat, Polish political reality has to be analyzed. The popularity of the party Law and Justice, which had been ruling in the years 2015 – 2019, was high before the 2019 election. All the polls guaranteed their victory. The public debate, in general, consisted of only two distinct groups of voters – the supporters of Law and Justice and the opponents. Due to the fact that the polls were clear in their conclusions, it can be assumed that each group had polarized expectations. Law and Justice supporters were, most possibly, waiting for the elections with the positive attitude, being sure that their party of choice is going to win. According to the psychological theorem, positive expectations regarding some event make people feel better. However, there is also the a side effect. When something positive is expected, the enthusiasm when it finally happens is not as vivid as it would be in the case of a surprise positive outcome. The inversed relation remains true for negative expectations (Golub et al., 2009). This raises the question whether such a mechanism could have made an impact. The hypothesis, which is tested when the sentiments are obtained, is:

*Auxiliary Hypothesis A:*

*Election do not have an impact on the average sentiment of tweets at the poviat level.*

The answer determines the choice of the model for explanation of voting behavior, which is introduced in Section 3. If the election do not have the impact, the sentiment in model can be treated jointly, with no split into sentiment before and after. However, if the hypothesis is rejected, two separate models will be needed.

Second issue with the data is associated with the location of the users – data representativeness problem. To investigate the share of tweets that contains users' locations, a simple test was conducted. During a few, randomly selected, days 100,000 tweets were collected. In each case the share of tweets with predefined location was about 1.5 %. That observation is consistent with other available statistics. They all confirm that the analyzed value is, indeed, really low (Haustein & Costas, 2015; Marciniec, 2017). The marginal share of users, who allow Twitter for "spying" on their location, raises a question whether such a group is representative. The representativity of Twitter data is the issue widely discussed in literature and has already been highlighted in Section 1 (Gayo-Avello et al., 2011). However, there is no way to overpass this without engaging users to the study with the help of, for example, surveys. Such a solution would highly downscale the population which can be analyzed.

The data representativeness doubts may also arise when the distribution of tweets around poviats is analyzed. Figure 4 presents how differentiated are the counts of tweets originating from specific poviats.

*Figure 4. Heatmap of count of tweets by poviat*

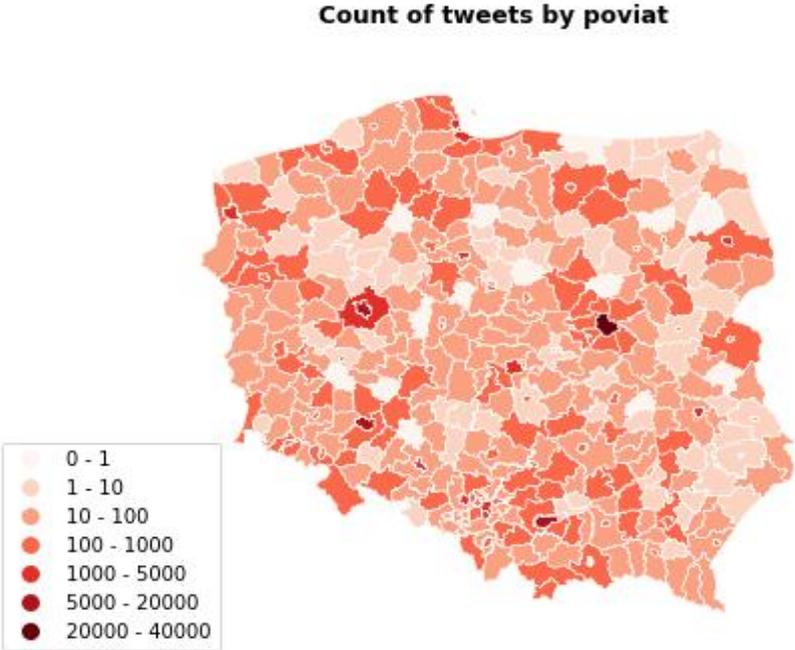

It can be easily seen that the counts of tweets are far from being evenly distributed. There are only a few places where the number of tweets is higher than 1000. They are visualized at Figure 5.

*Figure 5. Poviats with more than 1000 tweets*

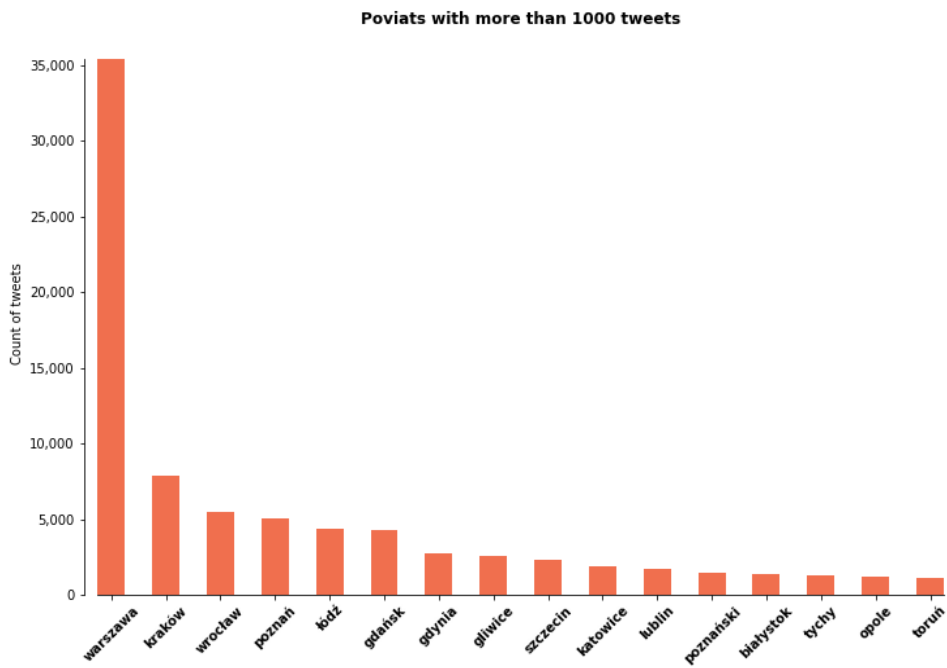

As it is consistent with the intuition and literature, all the places with high number of tweets are urban areas. 13 out of 16 poviats are, so called, cities at the poviat rights.[9] However, to get the full picture, count of tweets from a specific poviat should be weighted with the number of its inhabitants. This relation is visualized on Figure 6.

---

[9] City at poviat rights is a type of poviat regulated by Polish law. It is a city, which is in the same time a capital of poviat and a poviat itself.

*Figure 6. Population weighted count of tweets at the poviat level*

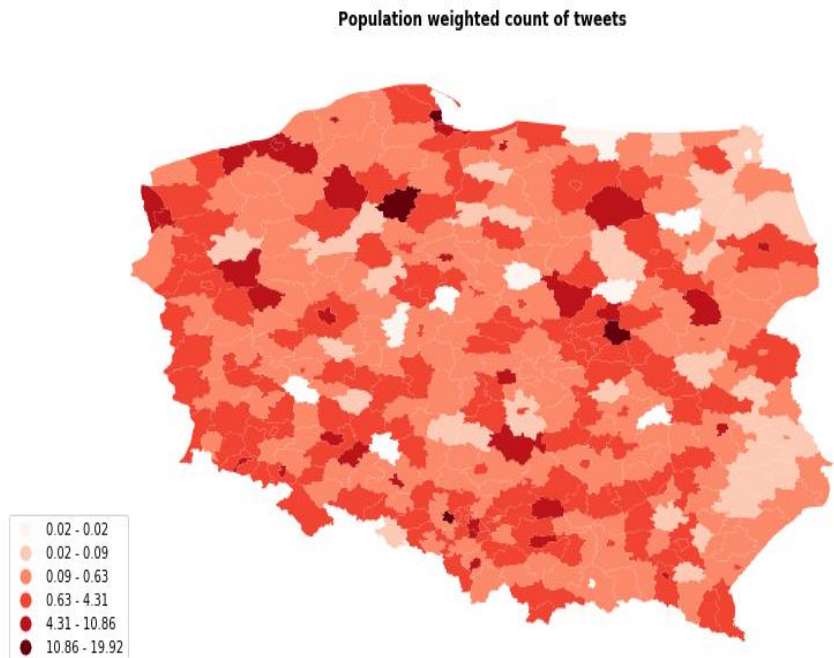

As it could have been expected, the weighted count of tweets is not balanced. The question is whether that is just how the reality looks like or it is some kind of "Twitter API" phenomenon.[10] To address that doubt simple OLS regression have been introduced with the count of tweets as the dependent variable and the number of poviat inhabitants as the explanatory one. Detailed results of the regression can be found in Annex 2. The t-test indicated that the number of poviat inhabitants is significant. Therefore, it is safe to state that the imbalance in the count of tweets is justified both by the literature and the statistical approach. It can be then assumed that it is just the reflection of the reality.

## 2.2. Two way approach to modelling – data preparation

There are two goals of sentiment analysis in this work. First, more obvious, to label the tweets data with sentiment scores in a best possible way. The second is to test different models, seeking the one with the best balance between accuracy and easiness of application. This strategy is additionally determined by the specificity of the data. The tweets are in Polish, which is not as widely supported in sentiment analysis as English language. Therefore, to get the advantage of already existing tools for the sentiment classification in English, the modelling in this work is carried out in two "ways":

- Modelling on tweets in its original version in Polish
- Modelling on tweets machine translated to English.

---

[10] "Please note that Twitter's search service and, by extension, the Search API is not meant to be an exhaustive source of Tweets. Not all Tweets will be indexed or made available via the search interface" (Twitter, 2020)

The structure of this section is determined by the duality above. First, it is crucial to discuss the whole conception of using machine translated texts in a sentiment analysis. Second, all the text preprocessing actions, which have been introduced, are talked over. To make the comparison between different sentiment analysis methods representative, the preprocessing is common both for original Polish tweets and these translated to English.

### 2.2.1. Tweets – machine translation

The vast majority of sentiment analysis tools is tailored to English. Pretrained libraries and word embeddings are mostly deployed for it. Most of the papers and discoveries also share this specificity. In the same time, it is clear that other languages are still widely used. It creates the need for the development of sentiment analysis tools also for non-English data.

One, straightforward, solution is to build models for the specific language. It is the first way approach proposed in this work. However, as it has been suggested by a few researchers, sentiment analysis of non-English texts can still be conducted with English based tools on the machine translations of original content.

Balahur and Turchi (2012) used opinion annotated sentences from English and machine translated them to: French, German and Spanish. Afterwards, they introduced SVM Classifier to conduct sentiment analysis. In the worst case possible 8 % drop of accuracy was observed between model based on machine translated content and the one in English.

Similar approach to Balahur and Turchi (2012) was suggested by Barhoumi et al., (2018) Authors conducted sentiment analysis on Arabic dataset which was machine translated to English. Logistic regression model was used. First on pre-trained Arabic dataset and the second on translations. Both the approaches resulted in almost identical accuracy.

The examples described in two previous paragraphs suggest that, to some extent, sentiment analysis of non-English texts might be conducted on their machine translations without significant loss in quality. However, it is important to indicate, that none of the papers considered tweets as their input data. Therefore, the comparison of the performance of Polish-based models and machine-translated-English is an additional value of this study, because of the unusual data. The translations have been done with the usage of Google Translate API[11] and Azure Translate API.[12]

### 2.2.2. Tweets – text preprocessing

Text preprocessing is a key step to perform good quality sentiment analysis. Tweets are the real challenge in this matter. Being restricted to only 280 characters, they push people to use some kind of the tweet-specific language. Their content is full of unique features which can be problematic during the analysis. In this work the preprocessing must also address the fact the tweets are supposed to be machine translated. Actually, the preprocessing methods suggested in this subsection are quite strict. Machine translation tools are expensive and they charge for every character translated. There was then a strong need to exclude not only the elements of tweets, which are the least informative for sentiment analysis, but also these, which can significantly worsen the quality of translation.

The first step introduced in text cleaning is to eliminate features and elements of tweets which do not contain any special information. These are:

- Removing links
- Removing mentions – Twitter user can mention another user with @user_name.
- Removing additional whitespaces
- Removing non word characters – punctuations, digits
- Removing few words tweets – to simplify dataset and get rid off these observations which are not rather likely to have any influence on the analysis the decision has been made to remove the shortest tweets, which contain no more than three words excluding conjunction.

It is important to indicate that the assumption about the lack of relevant information is true as far as the elements mentioned above are considered only as the word-features. For example, usernames are not really useful when they are perceived as a text. However, such a feature might be used as a basis for social network analysis, which can be utilized to improve the performance of sentiment classification.

---

[11] https://cloud.google.com/translate/docs
[12] https://azure.microsoft.com/pl-pl/services/cognitive-services/translator-text-api/

**2.2.2.1. Hashtags**

Hashtags are, in general, a way to identify a topic of a tweet. They have hyperlinks which allow an user to see all the tweets, which cover a similar field. The problem with hashtags in the area of sentiment analysis is to find a proper way to distinguish these which can carry some emotions with these which do not. For example #awesome indicates a positive nature of a tweet but #elections is just a topic. Additionally, some hashtag might only be a subject of a discussion, being at the same time consisted of words, which carry some sentiment. This can be understand with the examples of two most frequently occurring hashtags in our dataset: #Silni Razem (#Strong Together) and #Koalicja Obywatelska (#Civic Coalition). The latter is the name of second largest political party in Poland and the first is the slogan they used in their online campaign. Nevertheless, if the hashtags were considered as the words, they would be rated as highly positive, when they, actually, should not be. To conquer such a threat, first 10 most commonly occurring hashtags were analyzed. They are presented in Table 1.

*Table 1. Most frequently occurring hashtags*

| Position | Hashtag Content | Number of occurrences | % of hashtags |
| --- | --- | --- | --- |
| 1 | silnirazem | 964 | 1.71 |
| 2 | wybory2019 | 813 | 1.44 |
| 3 | koalicjaobywatelska | 616 | 1.09 |
| 4 | pis | 409 | 0.73 |
| 5 | plkpl | 393 | 0.7 |
| 6 | poland | 355 | 0.63 |
| 7 | warszawa | 317 | 0.56 |
| 8 | lewica | 288 | 0.51 |
| 9 | polska | 287 | 0.51 |
| 10 | elevenf1 | 279 | 0.49 |

Hashtags were present in about 10% of all tweets. Most of them are related to politics and 2019 parliamentary election. It indicates that Twitter is a social media source highly engaged in this topic. Even the most popular hashtags do not stand for a significant share in the total count.

Also, none of them have any sentiment attached. After the analysis, the decision has been made to remove all the hashtags from tweets.

### 2.2.2.2. Emoticons

The emoticons provide users with the ability to simply reflect the emotions without the need to express them with words. They are really important in the field of sentiment analysis, which can be proven with the following example:

- I hate you 🤤
- I hate you

There are big chances that the first sentence is positive, when the second one is more likely to be negative.

The role of emoticons in sentiment analysis is a widely discussed topic. Especially in the case of tweets their influence cannot be underestimated. Machine learning algorithms need training data to be fed with. At the same time, it is highly resource-consuming to obtain sentiment annotated dataset. To create it, the reasonable number of tweets has to be extracted and then labeled by manual annotators with sentiment scores. To simplify this process, some researchers proposed to use emoticons for automatic annotation of tweets (Pak & Paroubek, 2016; Zhao et al., 2012). They assumed that for tweets, due to their shortness, the emotional value of emoji attached to the text, can be considered as the emotional value of the whole tweet. However, as Wang and Castanon (2015) argued, such the approach is rather too good to be true. They proved that, indeed, emoticons are the important element of sentiment analysis, but not as the only one feature. According to researchers there are only a few emojis which carry a straightforward sentiment with themselves. Most of the emoticons are highly dependent on the context.

To deal with the emojis in this work, the decision was made to leave in a tweet content only these which can be easily distinguished as a positive or negative and which count stands for at least 1% share of total count of emoticons. As the result 16 emojis were selected. They are presented in Table 2.

Table 2. Most frequently occurring emojis

| Position | Emoji | % of emojis |
|---|---|---|
| 1 |  | 12.28 |
| 2 |  | 4.65 |
| 3 |  | 4.29 |
| 4 |  | 3.65 |
| 5 |  | 3.56 |
| 6 |  | 2.61 |
| 7 |  | 2.33 |
| 8 |  | 2.23 |
| 9 |  | 2.01 |
| 10 |  | 1.94 |
| 11 |  | 1.92 |
| 12 |  | 1.82 |
| 13 |  | 1.75 |
| 14 |  | 1.62 |
| 15 |  | 1.57 |
| 16 |  | 1.31 |

### 2.2.2.3. Spelling mistakes

Spelling mistakes are a common issue attached to internet-specific language. They are also not valuable for sentiment analysis. Additionally, they might visibly decrease the quality of a machine translation, what can be understood with the following example, which is based on one of the tweets from the data.

- Original content: *Opowiadałem ten koszmar na snapie i się poryvzalem*
- Manual translation: *I was telling about the nightmare on the snap (Snapchat) and started crying.*
- Deepl Translate: *I was telling that nightmare on the snap and I was kidnapped.*

- Google Translate: *I told this nightmare on snap and got over it.*

The quality of the translation is far from being satisfying. There is also a sentiment mismatch observable. When Deepl's translation is negative, Google Translate's is positive. The techniques which were introduced to address the issue are precisely described in Annex 3. All tweets which included misspelled words were excluded.

**2.2.2.4. Lemmatization and stop-words removal**

After obtaining the preprocessed version of the tweets, both for Polish and English, lemmatization and stop-words removal were applied to create additional sets for model training. The goal of these methods is to reduce the number of words present in dataset. It can simplify a learning process for the machine learning algorithms and possibly improve their performance. A more detailed description of both of the techniques can be found in Annex 4. The comparison of results between models based on original content vs lemmatized/stop-words-free content is then the additional value of this study, especially that it is conducted for two languages.

**2.3. Training data**

Every machine learning model needs a training data to learn on it. In the case of tweets such the data consists of a tweet content and a corresponding sentiment score. In this work the publicly-available set of sentiment annotated Polish tweets is used (Mozetič et al., 2016). It consists of 113,153 tweets, which distribution is presented in Figure 7.

*Figure 7. Distribution of sentiment in training data*

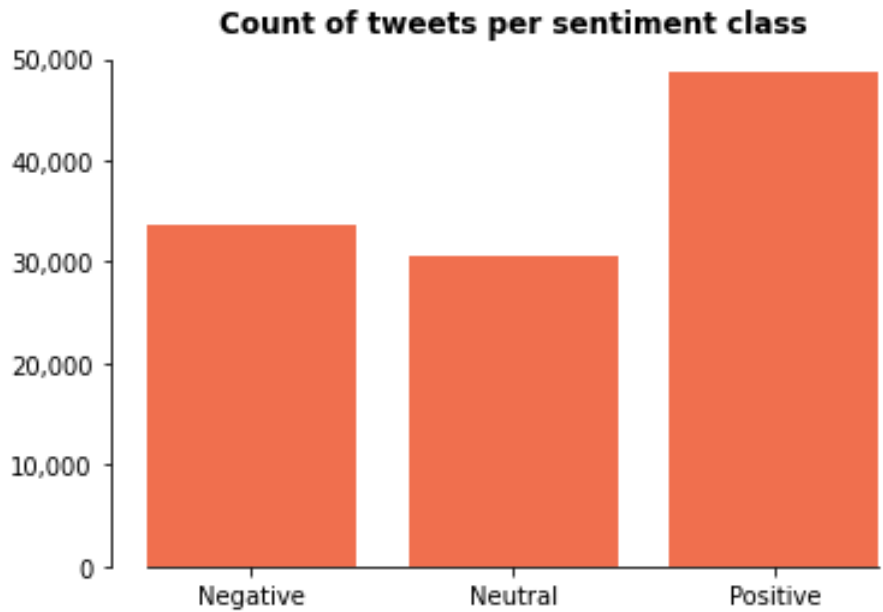

There is some positivity skewness in the data. However, it is not as explicit to signalize the data imbalance issue.

The tweets from the training data were the subject of the same preprocessing steps as these introduced for the data analyzed in this work. Additionally, because of the application of bilingual approach, they have been translated to English.

**2.4. Sentiment models**

The goal of the sentiment modelling in this work is to find models which would guarantee best balance between accuracy and easiness of deployment.

In the first wave two Python Libraries: Vader and Stanford NLP, which are able to handle three class sentiment analysis and are easy to use without any tuning, were tested. The limitation is, they are available only for English language. To address this issue, deep Convolutional Neural Network were constructed both for the classification of Polish and machine translated to English tweets. The architecture of the Network was inspired by the article of Cai (2019) and adjusted for three level classification. Cai conducted his analysis on English tweets, but focused only on negative and positive classes. The author achieved the best performance for his model in the combination with GloVe Embeddings (Pennington et al., 2014) trained on Twitter English Corpus[13]. The idea of GloVe – from Global Vectors is to mathematically capture relations between words which co-occur in the text corpus. When the embeddings are included in the sentiment modelling the algorithm is provided with additional knowledge. It can utilize not only the fact, it learns straight from the data, that some word, A, more often occurs in sentences labeled as negative, but also that word A is related to word B, which, again, leans toward negative sentiment. For Polish language, it is not easy to access high quality word embeddings. However, they can be trained inside the network, but what has to be said is that the quality of relations detected with such an approach is expected to be lower then for embeddings created with GloVe or word2vec (Mikolov et al., 2013) methods. Comparison of the performance, in terms of accuracy of the algorithms is presented in Table 3.

*Table 3. Classification accuracy for the proposed algorithms and different inputs*

| Model + Data | Accuracy in % |
|---|---|
| CNN + English Lemmatized + GloVe 100 | 55.3 |
| CNN + English Normal + GloVe 100 | 54.8 |
| CNN + Polish Lemmatized | 50.4 |
| CNN + Polish Normal | 49.7 |
| Stanford NLP + English Normal | 42.8 |
| Stanford NLP + English Lemmatized | 40.4 |
| Vader + English Lemmatized | 40 |
| Vader + English Normal | 32.1 |

The best accuracy was achieved by Convolutional Neural Network on lemmatized English words combined with GloVe 100 embeddings.

- CNN outperformed available libraries – Stanford and Vader.
- Vader, for English in its original form, got, actually, worse accuracy than random classifier for three-classes problem would.

---
[13] https://nlp.stanford.edu/projects/glove/

- Additionally, the lemmatization slightly improves classification for English, but not for Polish.

CNN occurred to be much better than the pretrained tools. The reason for that might be the tweets as the input data. They are a specific type of text. Both Vader and Stanford NLP were not directly tailored for tweets and, as it has been proven, failed to properly classify them. Nevertheless, 55 % accuracy for the best CNN is still far from being satisfying. The confusion matrix at Figure 8 puts some light on possible explanation.

*Figure 8. Confusion matrix for CNN + English lemmatized + GloVe 100*

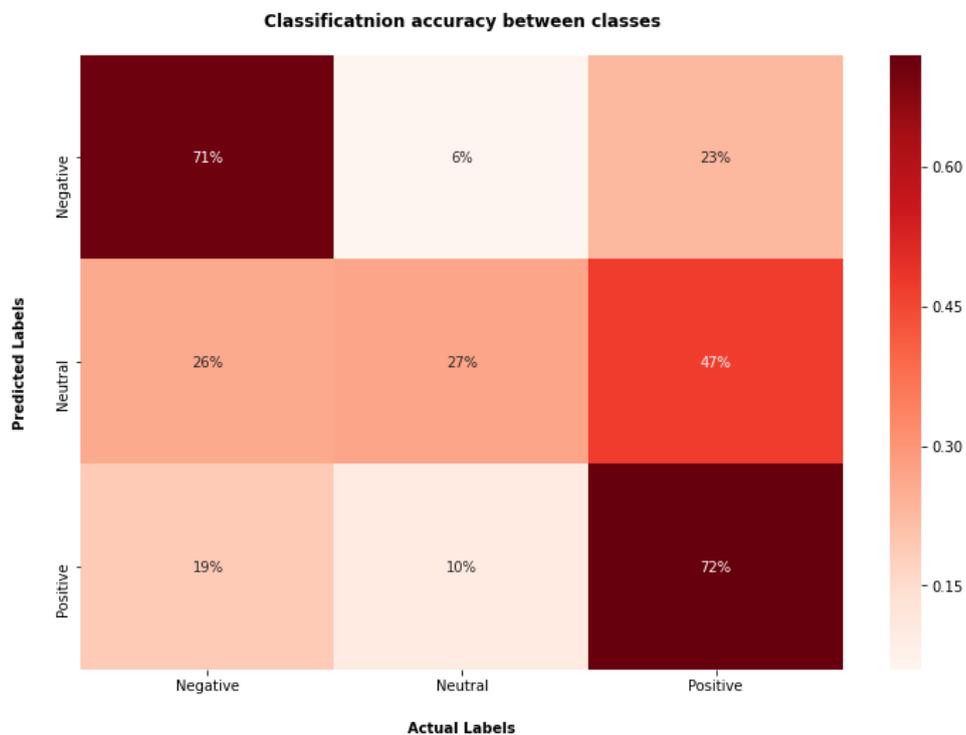

The algorithm fails mostly for the neutral tweets. The reason might be not in the model's inefficiency itself, but the input data being not properly annotated, which occurs to be especially problematic for the neutral class. To conquer this issue, another approach was suggested to switch from the three-class classification to the binary – neutral/positive. As CNN, based on the lemmatized English data, proved to be the best in previous section, it has been chosen for further binary analysis.

Before training the model for the binary classification, it was crucial to decide what should be the training data. The decision has been made to choose two strategies. First, straightforward, to use only the tweets from the training data, which are labeled as negative/positive. The second one was based on method called pseudo-labeling (Lee, 2013). The idea is simple. Train a model on labeled data, then predict values on a part of unlabeled data and combine datasets to train the final model. In other words, the final model is based on data which consists of actual values and predicted values, considered as actuals. In this work, the data considered as unlabeled would be one originally labeled as neutral. Basing on predictions already obtained, these tweets which had been labeled by the model as negative or positive were labeled as such, ones classified as neutrals were excluded.

Switching to the binary method opened also the possibility to use another pretrained tool, created by Huggingface[14], which utilizes BERT, the model from Transformers "family" which are now achieving best results in variety of NLP tasks.[15] At the end three models for the binary classification were introduced.

- CNN + Lemmatized English (Glove 100) + negative/positive/pseudo labels
- CNN + Lemmatized English (Glove 100) + negative/positive
- Pretrained Huggingface BERT

Additionally, to be sure of the quality of the classification, all the models were tested on the data of 26,354 labeled English tweets (Mozetič et al., 2016). The dataset is completely distinct and has no link to the training data. The accuracy of all the models is presented in Table 4. "Accuracy" is the accuracy the specific model achieved on the training data – how well it performed on the real (not pseudo) negative/positive labels. "Accuracy test" is the accuracy on the distinct test data.

*Table 4. Binary sentiment classification – accuracy comparison*

| Model + Data | Accuracy in % | Accuracy Test in % |
|---|---|---|
| CNN + English Lemmatized + Glove 100 (negative/positive/pseudo) | 81.7 | 51.6 |
| CNN + English Normal GloVe 100 negative/positive | 82 | 52.2 |
| Huggingface BERT | 66.7 | 71.1 |

---

[14] https://huggingface.co/transformers/usage.html
[15] https://paperswithcode.com/sota/sentiment-analysis-on-sst-2-binary

For the CNN model the obvious issue of overfitting has been encountered. Both CNN models achieved really good performance on the data they knew, but failed to work with the distinct dataset. Huggingface model occurred to be more stable for both datasets.

After all the modelling presented it is clear that sentiment analysis is a complicated task by itself. Additionally, tweets, due to their internet based language and short form, make the analysis even harder. The final model, which has been chosen to be one responsible for labeling the tweets collected for the purpose of this study, is the binary classifier – Huggingface BERT. It proved to be stable for two different datasets of tweets. However, even with this approach, it must be highlighted that the error of 30 – 40% of sentiment misclassified tweets has to be considered. The model will base on English lemmatized data, as it has shown to be the most informative. Another, valuable conclusion from the analysis conducted in this section is that the machine translation proved to be successful in the task. It can be then considered as the useful approach when working with sentiment analysis for non English tweets.

**2.5. Classification of Polish tweets – in search of regional happiness**

Huggingface BERT Sentiment Classifier, selected in Section 2.4, was used for the labeling of the tweets collected one month before and after 2019 Polish parliamentary elections. For the sake of the analysis, negative tweets are considered to have 0 value, positive: 1. Figure 9 presents the distribution of predicted sentiment of tweets.

*Figure 9. Predicted sentiment distribution*

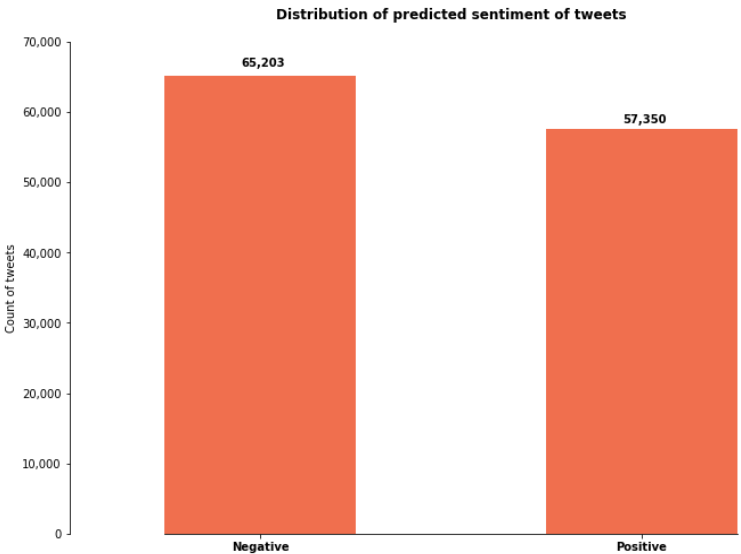

The sentiment of tweets is slightly skewed into the negative direction. However, clearly, the difference in the count between two classes is not large.

The key part of sentiment analysis conducted in this work is to find regional differentiation in sentiments between poviats. Average sentiment scores were calculated only for poviats with more than 100 tweets recorded. As the result of that assumption, the values were obtained for 126 poviats. The differences in scores can be seen in Figure 10.

*Figure 10. Average sentiment of tweets per poviat*

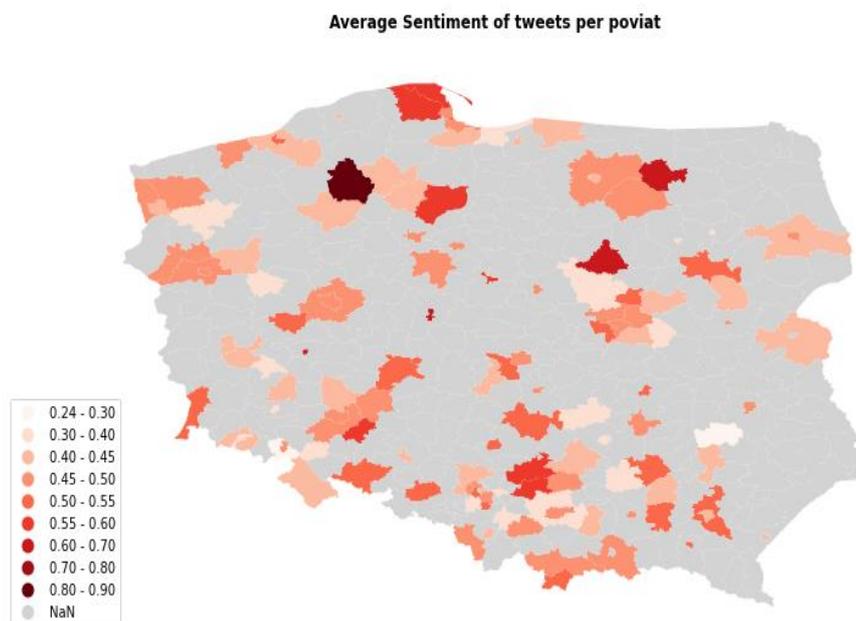

With the availability of the sentiment scores it is now possible to verify **Auxiliary Hypothesis A**: *Elections do not have an impact on the average sentiment of tweets at the poviat level*. The tweets were collected in the period one month before and after the elections. The questions is whether the sentiment of tweets from the month before the elections differ from this after the event. Two tests have been introduced for the sake of verification:

- OLS regression:
    - Y = Average Sentiment Scores per poviat from before and after the elections considered jointly.
    - X = Binary variable, which states whether the sentiment score is from before or after the elections.
- $X^2$-test for equality of proportions applied for each poviat separately.

For OLS regression t-test indicated that the explanatory variable is not significant. Detailed results of the regression are included in Annex 5. In other words, for the sentiment score it does not matter whether it was reported before or after the elections. In the case of $X^2$-test, the difference in proportion between periods before and after was insignificant for average values at the global level – $X^2(1) = 0.477, p = 0.490$. When tested for each poviat individually, it was significant only for 9 out of 126 poviats. This significance, however, was not associated with any specific characteristics of that poviats.

Both statistical approaches introduced indicate that the null hypothesis of ***Auxiliary Hypothesis A*** cannot be rejected. In other words, the election did not significantly influence the sentiment of the tweets at the poviat level. This conclusion has an important implication for Section 3 – the modelling can be done with the sentiment of all the tweets without separation for a period before and after the election.

**SECTION III**

**Determinants of voting behavior – regional happiness and socio-economic variables**

The poviat-level sentiment obtained in Section 2 is, by assumption, considered as the regional measure of happiness. In this section it is juxtaposed with socio-economic variables in order to explain the results of 2019 Polish parliamentary election. This section is divided into three parts. First, the political situation at the time of elections is described. Second, the choice of socio-economic data is explained. Third, the econometric model is introduced. At the end, three hypothesis, formulated in the closure of the Section 1, are verified.

**3.1. Polish parliamentary elections 2019**

The elections took place on 13th of October 2019. Law and Justice party, which had been ruling since 2015, was the favorite of the race. Figure 11 presents which party achieved the best score in specific poviats. The analysis is limited only to these poviats for which the sentiment score was obtained.

*Figure 11. Political party with the highest vote share for specific poviat*

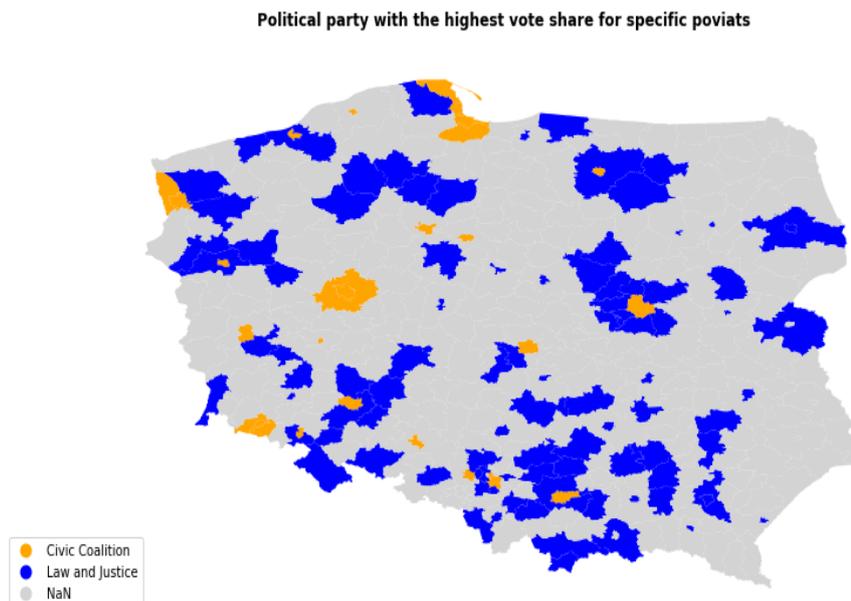

Only two political parties – Law and Justice and Civic Coalition won in selected poviats. The first one dominated most of the regions, having the highest vote share in 98 out of 126 poviats analyzed. The geographical pattern can be observed – the number of poviats, where Civic Coalition was the winner, increases to north-west direction. To confirm this observation the differentiation in vote share for Law and Justice is visualized on Figure 12.

*Figure 12. Vote share for Law and Justice*

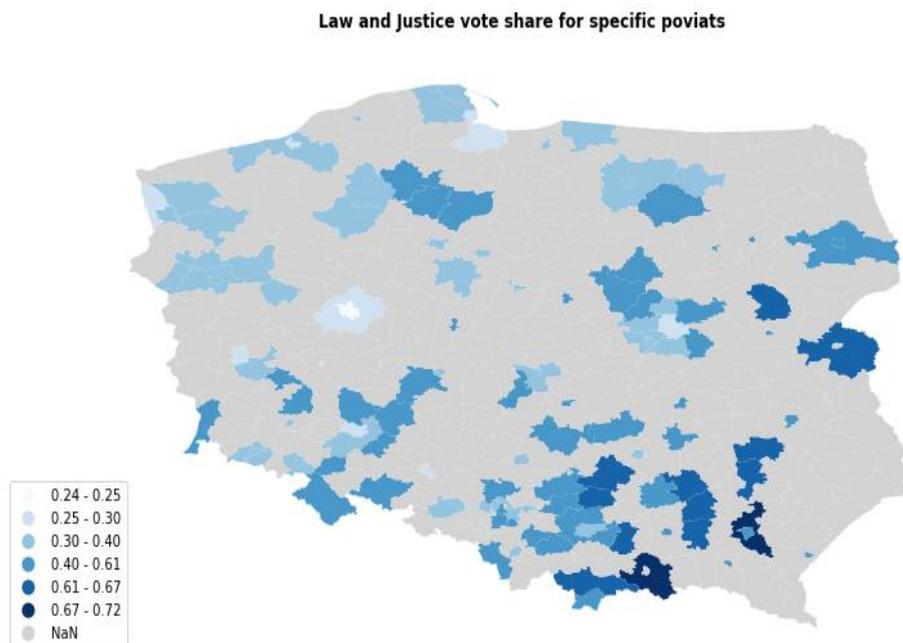

The popularity of Law and Justice is highest in south-eastern regions of Poland and decreases with the move into north-west direction. This remains consistent with the observation drawn from Figure 11. Geographical phenomenon, noticed, has implications for the model, which is introduced in Section 3.3. Due to the dominance of Law and Justice as a winning party in poviats, Law and Justice vote share is the dependent variable.

**3.2. Selection of socio-economic variables at the poviat level**

The selection of the variables is inspired by the list of features, which are the most likely to influence voting behavior, suggested by Fielding (1998) and already discussed in Section 1. All the socio-economic data incorporated to this study comes from Regional Data Bank[16] which is provided by Polish Central Statistical Office[17]. Some of the features proposed by Fielding had their direct match in the database:

- The unemployment rate
- The proportion of the population holding a higher educational qualification

---
[16] In Polish: Bank Danych Lokalnych https://bdl.stat.gov.pl/BDL/start

For the rest there was a need to find metrics which seem to be the suitable replacements. They are presented in Table 5.

*Table 5. Selection of Fielding-like variables*

| **Suggested by Fielding** | **Suggested Replacements from Regional Data Bank** |
|---|---|
| Average current disposable household income | Average gross monthly salary |
| Average household wealth | Balance of internal migrations per 1 citizen, |
| The proportion of the population employed in occupational classes 1 and 2 | Not available |
| The proportion of the working population employed in agriculture | Urbanization level |

Average current disposable household income could not have been obtained from Regional Data Bank at the poviat level. Average gross monthly salary has been chosen as the best possible replacement. However, it is clear that this substitution is not perfect and the variables are not the same. Also, average household wealth was not available as a straightforward variable. Studies suggest that there is an inversed relationship between wealth and migration, excluding developing countries, (Dustmann & Okatenko, 2014; Abramitzky et al., 2013). Therefore, Balance of internal migrations per 1 citizen was selected as the replacement. Urbanization level was picked as the variable which is able to map the proportion of the working population employed in agriculture. No measures were found which can reflect the proportion of the population employed in occupational classes 1 and 2 at the poviat level.

    Besides the variables proposed by Fielding (1998), to verify **Hypothesis 2** the model need to include sentiment, obtained in Section 2, perceived as the regional happiness measure. Additionally, to control for the conservativeness of poviats – some studies suggest that such

---

[17] In Polish: Główny Urząd Statystyczny

a variable might be useful to explain voting behavior (Powdthavee & Dolan, 2008) – the arbitrary choice was made and the Number of divorces per 1 citizen was selected. Finally, to control for the differentiation between less and more urban areas, binary variable[18] – City at the poviat right – was introduced. City at the poviat rights is a city which is at once a capital of a poviat and a poviat itself. The list of all the variables used to explain vote share of Law and Justice party in 2019 Polish parliamentary elections is presented below (the year the variables were recorded are in brackets).

- Unemployment rate (2018)
- Average gross monthly salary (2018)
- The proportion of the population holding a higher educational qualification (2011[19])
- The proportion of the population holding a medium educational qualification (2011)
- Balance of internal migrations per 1 citizen (2018)
- Urbanization level (2018)
- Median of age (2018)
- Number of divorces per 1 citizen (2018)
- Sentiment – average sentiment obtained in Section 2
- City at the poviat rights

**3.3. Models**

The variables described in Section 3.2. are the basis for the OLS regressions proposed to find voting behavior determinants and verify the hypothesis stated in Section 1. The first model – **General model –** introduced is based on all the possible variables. It is described with Formula 1.

---

[18] 1 indicates that this city is at the poviat rights.
[19] https://stat.gov.pl/spisy-powszechne/nsp-2011/

**Formula 1.** OLS regression with all variables included

*Law and Justice vote share$_p$ = 1+ Sentiment$_p$ + Urbanization level$_p$ +Number of divorces per 1 citizen$_p$ + Balance of internal migrations per 1 citizen$_p$ + Unemployment Rate$_p$ + Average gross monthly salary$_p$ + Median of age$_p$ + The proportion of the population holding a higher educational qualification$_p$ + The proportion of the population holding a medium educational qualification$_p$ + City at the poviat rights$_p$*

Where:

    *p* – represents data recorded for specific poviat.

The results of the regression for **General model**, presented in the second column of Table 6, were not fully satisfying, due to insignificance of half of the variables. In order to improve the performance, stepwise regression was introduced to choose the model with best AIC. Finally, new model – **Model 2**, described with Formula 2, was proposed.

**Formula 2.** OLS regression with best AIC

*Law and Justice vote share$_p$ = 1+ Sentiment$_p$ + Urbanization level$_p$ +Number of divorces per 1 citizen$_p$ + Balance of internal migrations per 1 citizen$_p$ + Median of age$_p$*

Results of **Model 2** are presented in the third column of Table 6.

*Table 6. OLS models comparison*

|  | **General Model** | **Model 2** |
|---|---|---|
| Intercept | 0.4246*** <br> (0.0070) | 0.4246*** <br> (0.0070) |
| Sentiment | -0.0144** <br> (0.0070) | -0.0133* <br> (0.0070) |
| Urbanization level | -0.0526*** <br> (0.0170) | -0.0439*** <br> (0.0140) |
| Number of divorces per 1 citizen | -0.0261*** <br> (0.0090) | -0.0278*** <br> (0.0090) |
| Balance of internal migrations per 1 citizen | -0.0503*** <br> (0.0110) | -0.0459*** <br> (0.0080) |
| Unemployment rate | -0.0078 <br> (0.0090) |  |
| Average gross monthly salary | -0.0042 <br> (0.0090) |  |
| Median of age | -0.0199** <br> (0.0090) | -0.0208** <br> (0.0090) |
| The proportion of the population holding a higher educational qualification | 0.0038 <br> (0.0130) |  |
| The proportion of the population holding a medium educational qualification | -0.0056 <br> (0.0090) |  |
| City at the poviat rights | 0.0090 <br> (0.0300) |  |
| R-squared | **0.5170** | **0.5100** |
| Adjusted R-squared | 0.4740 | 0.4900 |
| Prob (F-statistic) | <0.0000 | <0.0000 |
| AIC | -272.2 | -280.6 |

Standard errors are reported in parentheses. *, **, *** indicates significance at the 90%, 95%, and 99% level, respectively.

The most important, considering the goal of this thesis, is the significance of the sentiment variable. It is also interesting to analyze its coefficient, which indicates that the happier people are the less likely they are to vote for Law and Justice. This observation is not intuitive. One would rather expect that if people are happy with their lives, they are also satisfied with a current government. There is then no reason for them to vote for another parties. However, it might be the case that people who vote for Law and Justice are, generally, less satisfied with their living. Future studies in this topic might bring interesting insights into the differentiation of life satisfaction between Polish citizens.

Law and Justice is vividly more popular in not wealth rural/small-city areas among conservative voters. This is indicated by the signs of the coefficients of other significant explanatory variables of the model. They all seem to follow the expected behavior.

- The higher the urbanization, the worse the vote share for Law and Justice. This is the reflection of reality, in which the ruling party is more popular in non-urban areas.
- The more positive the balance of internal migrations, the lower the vote share for Law and Justice: This, again, remains consistent with the real-life. The highest balance is observed in wealthy urban-areas, which, again, are not a main political basis for the ruling party.
- The higher the number of divorces (the less conservative the poviat) the lower is the vote share for Law and Justice, which is expected, due to the conservative nature of the ruling party.

The only one unexpected estimation has been obtained for Median of age. Figure 13 visualizes what was the distribution of the vote share for Law and Justice among different age groups in 2019 election.

*Figure 13. Vote share of Law and Justice among specific age groups*

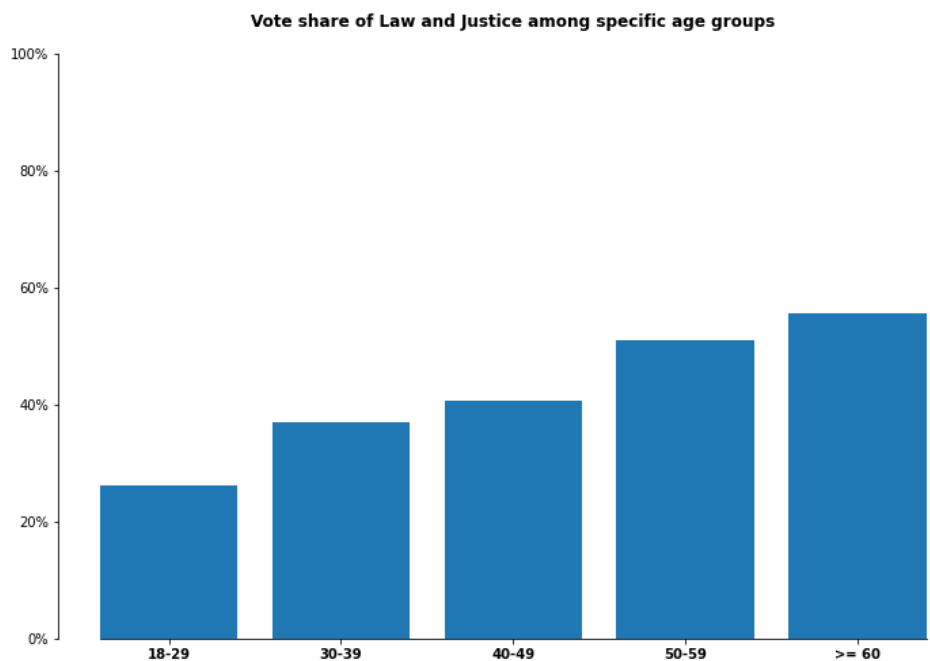

*Source:* Own elaboration based on https://wiadomosci.gazeta.pl/wiadomosci/7,143907,25241642,wyniki-exit-poll-frekwencja-wedlug-wieku-jak-glosowaly-konkretne.html

With the Figure 13 it is clear that the coefficient obtained for Median of age is counterintuitive. That issue might have its source in the variable not being a proper reflection of the age structure of specific poviats. In future studies other approach to control for age should be applied – for example: the number of people in the oldest age group in a poviat divided by the total number of people in a poviat.

The insignificance of the variables from **General model** should also be discussed. The unemployment rate and the variable controlling for an income – average salary – were both significant in another study, which was focused on the explanation of voting behavior at the regional level in Poland (Grabowski, 2018). The author analyzed 2015 Polish parliamentary election. There is then a question why mentioned variables occurred to be not significant in this work. The possible explanation might come from the comparison of results Law and Justice obtained in 2015 and 2019 among different groups of voters, clustered by age, education, social status. In 2015 Law and Justice dominated the whole political scene. The party won even in the largest cities and among the best educated people, which had not been seen as the basis of their political electorate for the long time ("*Wyniki wyborów. PiS pierwszą siłą. Wśród młodych, wykształconych, z miast…*", 2015). In 2019, Law and Justice was also the winner among most of the groups of voters. However, they lost in a couple – largest cities, best educated people. (Zaborowska, 2019). Therefore, the relations in case of this study were not as straightforward as in the case of 2015 parliamentary election. What should be also pointed out is the insignificance of the educational variables and the one controlling for larger urban areas – City at the poviat rights, combined with not intuitive coefficient of median age. This observation is at odds with a popular concept in Polish public debate about "young, well-educated living in large cities", being the electorate of liberal parties opposed to Law and Justice (Władyka & Janicki, 2012). This conception is quite old and the voting behavior of this group might have changed. It would be beneficial to verify the truthfulness of this observation in future studies.

**3.4. Hypotheses verification**

Since the models and their estimations have been discussed, it is possible to address the hypotheses from Section 1.

*Hypothesis 1:*

*Results of the elections at the poviat level can be explained with socio-economic variables, as suggested by Fielding (1998).*

**General model** and its results, presented in Table 6, show that only two of the variables, suggested by Fielding as valuable predictors of elections, are statistically significant – Urbanization level[20] and Balance of internal migrations[21]. Non-zero impact on the dependent variable can also be observed for a few other socio-economic features, which have not been mentioned by Fielding. Therefore, it can be concluded that the election results can be explained with selected socio-economic characteristics. However, the choice of these explanatory variables should be tailored to the analyzed country.

*Hypothesis 2:*

*Sentiment of tweets, aggregated at the poviat level, is a statistically significant predictor of election outcomes.*

The sentiment of tweets was found to be statistically significant as the explanatory variable of election outcomes, both for the **General model** and **Model 2**. This finding might be an additional value for vast political and social studies. All the variables, excluding sentiment, used for the vote share explanatory model in this work, evolve slowly in time. It is then difficult to exploit them to predict outcomes of future elections. Sentiment of tweets, however, is the picture of current state of moods in specific regions. Taking all this into consideration, **Hypothesis 2** cannot be rejected.

---

[20] Variable suggested by Fielding was: The proportion of the working population employed in agriculture, Urbanization level has been proposed as the replacement.
[21] Variable suggested by Fielding was: Average household wealth, Balance of internal migrations has been proposed as the replacement.

**SECTION IV**

**Summary and conclusions**

The aim of this research was to utilize the information of sentiments of the tweets and combine it with socio-economic variables to explain voting behavior at the poviat level in Poland. This required collecting and conducting an analysis of tweets, observed in the period around 2019 Polish parliamentary election.

The sentiment analysis was a complicated, multistage modelling process. It was conducted both on the tweets in their original language – Polish and on English machine-translations. The methods proposed varied from using pretrained tools – Vader, Stanford NLP[22], Huggingface BERT[23] – all being available only for input data in English, to the models created from scratch to work with both languages analyzed. At the beginning, three-class classification was tested. However, the performance of the algorithms, reported in Table 3, was not satisfying. The best, 55.3 % accuracy was obtained by Convolutional Neural Network on the lemmatized, machine-translated to English tweets, with 100 dimensional GloVe embeddings. The CNN outperformed both Vader and Stanford NLP, which might be the indication, that these libraries are not suitable for sentiment analysis of tweets. The important lesson from this stage of the modelling was that English machine-translations worked better with the models proposed than the tweets in original – Polish language. Nevertheless, 55.3% accuracy was not enough. To address that issue, the decision has been made to limit the sentiment analysis to the binary classification. In this task only the tweets machine-translated to English and lemmatized were used. Three models were tested: Binary CNN, Binary CNN with pseudo-labeling and pretrained – Huggingface BERT. In the matter of accuracy and stability, Huggingface BERT model outperformed the other two. It had almost the same performance on two, completely distinct datasets of tweets, achieving 66.7 % and 71.1 % accuracy. The classification quality was still not fully satisfying, especially that some loss of information might have occurred due to the switch from the three-class to the binary classification. However, due to the lack of prospects for the improvement, Huggingface BERT was selected as the model to predict sentiment scores for the tweets which were collected for the purpose of this study.

With the tweets labeled with sentiment scores it was possible to address the main goal of the work – explaining the voting behavior at the poviat level in Poland and verifying the two research hypotheses, related to predicting election outcomes with sentiment of tweets and socio-demographic variables. The OLS regression models have been utilized to verify the research hypotheses. Law and Justice vote share in poviats was selected as the dependent variable. Average sentiment along with selected socio-economic features in line with the literature (Fielding, 1998), were employed as explanatory features.

---

[22] Suitable for three-class classification
[23] Binary classification

The sentiment of the tweets was found to be a significant predictor of voting behavior. The observed relationship was negative, indicating that the more positive the tweets in a particular poviat are, the lower the expected election result for the ruling party – Law and Justice. One possible explanation is that the Law and Justice receives more votes in regions that are relatively unhappy, however, due to cross-sectional character of the data it is not possible to verify this stipulation. In the case of the other socio-economic variables, only a few of them proved to be significant. From the variables proposed by Fielding (1998) a non-zero impact on the dependent variable was observed for: Urbanization level and Balance of internal migrations per 1 citizen. In both cases the coefficients had expected signs. Besides these two features, Number of divorces per 1 citizen and Median of age proved to be significant. The former was selected to control for the conservativeness of the region, since Law and Justice is a conservative party. The sign of the coefficient indicated, as expected, that the lower the conservativeness of the region the lower is the vote share for the ruling party. To sum up this part – both main hypotheses could not be rejected. This means that for voting behavior as the dependent variable, socio-economic features and sentiment of tweets proved to be useful as the predictors.

Overall, the main contributions of this study can be presented together on the following list:

- Sentiment of tweets proved to be an informative predictor for voting behavior. This is the most important conclusion coming from this work. Social medias and tools for sentiment analysis are in constant development. It means, that both the availability of information and the ability of algorithms to process them, increase. The nature of social medias make them an always up-to-date source of data. It can be then expected, that, in the future, the approaches which utilize sentiment analysis of social medias will be able to replace or, at least, improve standard polls.
- Sentiment analysis conducted on machine-translations do not bring significant decrease in the performance of the algorithms. It is the important finding especially from a business perspective. A company, which would like to utilize sentiment analysis for some application, would not have to build it from scratch, because they could benefit from tools pre-trained for English. They would only have to invest in some API translation service.

- Socio-economic variables are useful in the explanation of voting behavior. However, their choice seem to be, at least to some extent, country specific. In other words, the variables which are suitable for one country, might not occur useful for the analysis conducted for another one.

- It is hard to measure the conservativeness level of some society. There is no straightforward variable. Therefore, the next valuable contributions of this work is that the probable replacement might be suggested – Number of divorces per 1 citizen. The variable proved to be both significant and intuitive in terms of the sign of the coefficient. Obviously, there is a question whether this conclusion is universal and would be applicable for another studies. It could be verified in future researches.

This study have several limitations that have to be acknowledged. The most important are related to the sentiment classification. The algorithm, which has achieved the best performance on the testing data and was chosen for the task, had still moderate (60-70%) accuracy. It can be expected that about a quarter of the tweets analyzed was misclassified. Additionally, the sentiment classification has been conducted on a binary (negative/positive) level. The neutral class was not included. The problem with neutral sentiment class is that its interpretation might be ambiguous. It can be just a label attached when the algorithm is not able to state whether a text is positive or negative. However, it can also indicate meaningful neutrality of a text, which would mean that the writer of the message is indifferent to a topic. With the exclusion of the neutral class in this work such uncertainty has been eliminated but, at the same time, it might be the reason for the lose of information. Another limitation of the study is the relatively modest selection of socio-economic characteristics. The variables included in the analysis were based on theoretical predictions of Fielding (1998), whose study was conducted on Scottish population. It would be beneficial to identify a variety of characteristics, which are specific for a Polish political culture. It is also important to remember about the widely discussed issue in the literature regarding Twitter analysis in general. Its population is not a representative reflection of the society. Future studies could extend this work and address all these limitations.

In summary, this study covers an important and future-oriented topic of the social phenomena in the age of social medias. With the constant development of technologies, it can be assumed that our "offline" lives will get more and more bounded with the "online world".

Even today we can observe that the social changes can be motivated and directed with the usage of social medias and such influence is likely to rise. That is why it is so important to create the methods to understand the online reality as well as possible. This thesis offers another methodological and empirical step towards this goal.

# LIST OF INDEXES

**Index of tables**

**Index of figures**

**Index of annexes**

# ANNEXES

**Annex 1. Obtaining desired data about a location from a tweet**

Based on OpenStreetMaps API (OpenStreetMap, 2020), the communes and voivodeships for cities, provided as the location in tweets, were obtained. The poviats, however, were not available from this source. Additionally, what had to be addressed was that there might be a few cities with the same name. The assumption was made to select always the city with the highest value of importance[24], which is the part of OpenStreet Maps API output. Such an approach might be responsible for introducing some error. However, places which are considered as more "important" are cities with the higher number of inhabitants. It remains then consistent with the fact that people from urban areas are more likely to be engaged in social media activity. To extract poviats, based on communes and voivodeships, Wikipedia scrapper, which had been written specifically for this purpose, was used.

**Annex 2. OLS regression – count of tweets from a specific poviat vs number of its inhabitants**

*Dependent Variable: Count of tweets from a specific poviat*

|  | **Coefficient** |
| --- | --- |
| Intercept | -1160.9806*** |
|  | (63.0150) |
| Number of inhabitants of specific poviat | 0.0145*** |
|  | (0.0000) |
| **R-squared** | 0.7900 |
| **Adjusted R-squared** | 0.7890 |
| **Prob (F-statistic)** | <0.0000 |

Standard errors are reported in parentheses. *, **, *** indicates significance at the 90%, 95%, and 99% level, respectively.

---

[24] "The major weight of importance comes indeed from the Wikipedia link count. If no article can be found for an object, the base score is based on the object rank." (Hoffmann, 2013)

**Annex 3. Spelling-mistakes**

To address this issue, probability based spelling correction (Norvig, 2016) was first tested. However, when the author proved its high efficiency with English language, with the accuracy about 80% – 90%, in the case of Polish language the method was not successful. Another strategy introduced, utilized a publicly available dictionary. For each tweet it was verified whether used words are also present in the dictionary. If at least one word was not, the tweet was labeled as incorrect. With this approach 4.7 % of "misspelled" tweets were found. Taking into consideration that misspellings can have a really bad influence on a quality of a translation and the fact that the share of tweets containing misspells is low, the decision was made to remove these tweets from dataset.

**Annex 4. Lemmatization and stop-words removal description**

Lemmatization allows for obtaining, so called, lemma for multiple versions of the same word. It can be easily understood with the following example: "(…) the verb 'to walk' may appear as 'walk', 'walked', 'walks' or 'walking'. The base form, 'walk', that one might look up in a dictionary, is called the lemma for the word"[25]. Lemmatization was done with Python Stanza library both for Polish and English.

Stop words are words which occur most frequently for specific language, and generally do not carry any valuable information by themselves. Therefore, it is a common practice in different NLP fields to exclude them. However, in the case of sentiment analysis there is some risk which is bounded to this approach, which can be represented with example below:

- Original sentence: He is not bad
- Lemmatized sentence: bad

When the first sentence is rather neutral/positive, the second is negative. It proves that, in some cases, stop words, even if they do not contain any sentiment value alone, influence the sentiment when analyzed jointly with other words present in sentence.

---

[25] https://en.wikipedia.org/wiki/Lemmatisation

**Annex 5. OLS regression – significance of sentiment before and after the 2019 Polish parliamentary election**

|  | Coefficient | P-value |
|---|---|---|
| Intercept | 0.4724 | 0.000 *** |
|  | (0.008) |  |
| Binary Flag – Sentiment | -0.0061 | 0.578 |
| Before and After Election | (0.011) |  |
| **R-squared** | 0.001 |  |
| **Adj. R-squared** | -0.003 |  |
| **Prob (F-statistic)** | 0.578 |  |

Standard errors are reported in parentheses. *, **, *** indicates significance at the 90%, 95%, and 99% level, respectively.